\documentclass{rspublic}

\usepackage[dvips]{graphicx}
\usepackage{amssymb,amsfonts,amsmath}

\def\lsim{\mathrel{\rlap{\lower4pt\hbox{\hskip1pt$\sim$}}
    \raise1pt\hbox{$<$}}}                
\def\gsim{\mathrel{\rlap{\lower4pt\hbox{\hskip1pt$\sim$}}
    \raise1pt\hbox{$>$}}}                
    
\def\be{\begin{equation}}    
\def\ee{\end{equation}}    

\def\({\left(}
\def\){\right)}
\def\[{\left[}
\def\]{\right]}

\def\a{\alpha}
\def\b{\beta}
\def\f#1#2{\frac{#1}{#2}}

\def\de{\delta}

\def\del{\nabla}

\def\p{\pi}
\def\r{\rho}

\def\<{\langle}
\def\>{\rangle}

\providecommand{\abs}[1]{\lvert#1\rvert} 

\begin{document}

\title[Modifying Gravity: You Can't Always Get What You Want]{Modifying Gravity: You Can't Always Get What You Want}

\author[G.D.~Starkman]{Glenn D.~Starkman$^1$}

\affiliation{$^1$ Dept of Physics, Center for Education and Research in Cosmology and Astrophysics,
and Institute for the Science of Origins, Case Western Reserve University, Cleveland OH, USA 44106}

\label{firstpage}

\maketitle

\begin{abstract}{modified gravity, dark matter, dark energy}
The combination of General Relativity and the Standard Model of particle physics disagree with numerous
observations on scales from our Solar System up.  
In the canoncial concordance model of $\Lambda{CDM}$ cosmology,  many of these contradictions between
theory and data are removed or alleviated by the introduction of three completely independent new components of
stress-energy -- the inflaton, dark matter, and dark energy. Each of these in its turn is meant to have (or to currently) 
dominate the dynamics of the universe.   There is, until now, no non-gravitational evidence for any of these dark sectors;
nor is there evidence (though there may be motivation) for the required extension of the standard model.  
An alternative is to imagine that it is General Relativity that must be modified to account for some or all of these
disagreements.  Certain coincidences of scale even suggest that one might expect not to have to make independent
modifications of the theory to replace each of the three dark sectors.     
Because they must address the most different types of data, 
attempts to replace dark matter with modified gravity are the most controversial.  A phenomenological model 
(or family of models), Modified Newtonian Dynamics, has, over the last few years seen several covariant realizations.
We discuss a number of challenges that any model that seeks to replace dark matter with modified gravity must face:
the loss of Birkhoff's Theorem, and the calculational simplifications it implies; the failure to explain clusters,
whether static or interacting, and the consequent need to introduce dark matter of some form, whether hot dark 
matter neutrinos, or dark fields that arise in new sectors of the modified gravity theory; the intrusion of cosmological
expansion into the modified force law, that arises precisely because of the coincidence in scale between 
the centripetal acceleration at which Newtonian gravity fails in galaxies, and the cosmic acceleration.  We conclude
with the observation that, although modified gravity may indeed manage to replace dark matter, it is likely to do 
so by becoming or at least incorporating, a dark matter theory itself.
\end{abstract}

\section{Introduction}

The Standard Model of particle physics, augmented by the generally benign addition of a mechanism for generating neutrino flavor mixing, successfully describes a remarkable range of short distance phenomena.  From the highest energy collisions at particle colliders (and in cosmic rays)
to the internal structure of stars,  observations match this model's predictions to the precision 
with which physicists have been able to calculate them.    Meanwhile from the scale of the Earth's
orbit around the sun down to just a few microns, General Relativity accurately describes all known 
gravitational phenomena.

Much has been made of our inability to reconcile quantum field theory -- the mathematical and
conceptual basis of the Standard Model -- with general relativity on ultra-short scales.  Addressing
this failure is the primary motivation behind string theory, loop quantum gravity, and other approaches
to quantum gravity.   However, while the ultra-violet failings of the standard model and GR are
as yet theoretical -- no experiment currently probes the quantum gravity regime -- 
 the infrared failings of this pair are real.  {\bf At all scales above that of the solar system, the combination of
 General Relativity and the Standard Model contradict observations.}
 
Viewing the past century of astronomical and cosmological data as a massive
assault on our canonical theories is a non-traditional perspective, but one that
underlies the whole program of modified gravity.  It is therefore appropriate to 
recall what a standard model general relativistic cosmology would look like and
review the ways in which observations contradict its predictions.   This we shall
do in section one, where we shall also review the standard "dark side" solutions to these problems
and see how it leads to a concordance cosmology that would have made William of Ockham shudder.

In section two, we will  present the possibility  of modified gravity as a solution to the
failure of general relativistic standard model cosmology, focusing on frameworks
where it offers a comprehensive solution to both the missing mass in cosmic structures
(aka the missing or dark matter problem) and the observation of cosmic acceleration
(aka the dark energy problem).  

In the third section, we will discuss the challenges these comprehensive modified gravity
theories face, and why they seem doomed to give Father William even less comfort
than their dark competitors.

\section{The many failures of Standard Model General Relativistic Cosmology}

The standard model contains a very limited number of stable ingredients from which to assemble
the universe we observe.  Protons and neutrons (and their anti-particles) are the only stable baryons.
They can be combined into just 255 known stable nuclei, ranging from hydrogen to lead-208,
however, the universe contains only trace amounts of anything other than hydrogen ($^1$H
in particular) and Helium-4.  (Deuterium ($^2$H) has an abundance of ~$0.01\%$  relative to $^1$H,
and Helium-3 an abundance of a few $10^{-5}$.  Among higher atomic number elements, oxygen
has the greatest abundance, but is still less than $0.1\%$.)   This paucity of heavy elements should come as something
of a surprise since many nuclei are more stable than hydrogen or helium.  It results from a variety
of accidents of nuclear physics and cosmology.  These include the combination of the small binding 
energy of deuterium and the large number of photons per baryon in the universe which result in 
the so-called deuterium bottleneck -- the delay in the synthesis of elements heavier than hydrogen
until the universe has cooled to a temperature where characteristic energies are far far below the 
characteristic activation energies of nuclear fusion processes.  The absence of any stable isotope of
atomic mass number eight, further inhibits processing of anything beyond lithium and beryllium.
All heavier isotopes are produced almost exclusively much later - in stars and in interstellar cosmic ray
interactions. Meanwhile, the only stable negatively 
charged particle that can neutralize the charge of a nucleus to form an atom is an electron. 

From the point of view of a cosmologist, this  parsimony is fortunate, because it translates into a rather
limited number of possible stable objects of any size.  Basically, there is nothing between atoms bound together
by electromagnetism (and consisting almost entirely of nuclei bound by strong interactions) and 
Jupiter-sized brown dwarfs that can play a significant role in cosmology, the one exception being primordial
black holes which, above a mass of about $10^{10}$kg, would evaporate (Hawking 1974) sufficiently slowly to play
an important dynamical role in the universe.

We can now proceed to describe some of the failures of a purely standard model early universe based solely 
on general relativity (SMGRC).

Let us start with the implications of the relatively simple process of nucleosynthesis in the early universe
alluded to above.
This process, which takes place when the universe is just a few minutes old in the SMGRC
(i.e. a few minutes after it "emerges" from the Big Bang singularity),  results in the binding of about
one quarter of all baryons into helium-4 nuclei, and the production of trace amounts of deuterium,
helium-3, and lithium-7.   Relatively simple numerical calculations (Alpher 1948; 
Peebles 1966ab, Wagoner 1967) show that the relative abundances of these isotopes 
are sensitive functions of the number density of baryons
(compared to say the number density of photons). These isotopes can be identified spectroscopically 
in many objects in the contemporary universe -- stars, HII regions, ... Although the current abundances have been
changed to greater or lesser extents in different systems from the original primordial abundances, 
by chains of inference of varying elaborateness and reliability, these primordial abundances can be inferred.
A great success of SMGRC is that these several primordial abundances are mutually consistent, even 
though there is only one free parameter of much influence -- the baryon-to-photon ratio.   This turns out to be
approximately $10^{-10}$.

Where success turns
to disaster is when we combine the inferred baryon-to-photon ratio with the current photon density 
(the number density of cosmic microwave background (CMB) photons, to be specific) which we can measure
with sensitive microwave detectors (Penzias 1965).  This allows us to infer the current number
density of baryons, and hence the current density of Standard Model matter,
\be\rho_{SM} \simeq 5\times 10^{-31} {\rm g}/{\rm cm}^3 . \ee
The difficulty comes when this is compared to the total energy density when inferred from the rate
of expansion of the universe using the Freedmann equation:
\be
\label{eqn:Friedmann}
H^2 = \frac{8 \pi G_N \rho}{3}.
\ee
where $H$ is the Hubble parameter -- the logarithmic time derivative of the rate of the cosmic scale factor $a(t)$.
Since today $H=H_0 \simeq 70 {\rm km}/{\rm s}/{\rm MpC}$, one finds that
\be
\rho = \rho_c \equiv \frac{3 H_0^2}{8\pi G_N}  \simeq 1 \times 10^{-29} {\rm g}/{\rm cm}^3.
\ee
We thus see that $\Omega_b \equiv \rho_{SM}/\rho_c=0.05$.
Within the context of SMGRC this discrepancy must mean that the geometry of the universe is strongly negatively 
curved since actually 
\be
H_0^2 = \frac{8 \pi G_N \rho}{3} + \frac{k}{a(t)^2}.
\ee
Unfortunately, detailed observations of the power spectrum of fluctuations in the CMB (Lange 2001, Balbi 2000, Spergel 2003)
set an upper limit on  $\Omega_{curv} = 1-\rho_{SM}/\rho_c \lsim 0.05$ which is significantly smaller than the inferred 
value in SMGRC of $0.95$.

The only possibility within the SMGRC of resolving this missing energy-density problem
 is that the other 95\% of the energy density of the universe is in the form of so-called 
primordial black holes (PBH) -- in particular black holes formed before nucleosynthesis.   
There is no mechanism within the SMGRC for the formation of long-lived black holes.  
That requires a first order phase transition late enough after the singularity that the mass contained within
the horizon can form a black hole with a lifetime to Hawking evaporation greater than the current age
of the universe -- about 10 billion years old.  The only Standard Model phase transitions that  qualify are the
electroweak phase transition, the colour-confinement phase transition and the chiral symmetry breaking phase transition.
But none of them are first order within the standard model.   Furthermore, the universe can be shown (Carr 2010) not to contain black holes with twenty times the average mass density as the matter,  indeed $\Omega_{PBH}\equiv\rho_{PBH}/\rho_c$ is constrained for $M_{PBH}$ to be less than (at most) $10^{-4}$.   Only for PBHs with mass less than or equal to the
Planck mass is this limit inapplicable, but that is precisely because they are in the ultra-violet regime where GR meets
quantum mechanics, and so the behaviour of such PBHs is ill determined -- in other words, they are outside the SMGRC.
Furthermore, as we shall discuss below, having all of that 95\% of the missing energy-density be of a form that
scales with redshift ($z$) as $(1+z)^3$ (as PBHs do) is not viable.  Most of it needs to depend much more weakly on $z$.

This was, of course, not the first case made for missing matter.  The original inference that the 
matter that we see is not sufficient to source the gravitational forces that we infer from 
observations dates back to Zwicky (1933) and his observations of clusters.   Indeed, in
all gravitationally bound systems from the length scale of dwarf galaxies up (ie. galaxies, groups and clusters)
one observes that (except in the systems' central regions)
the velocities of the objects that are apparently bound to the system exceed the local escape velocity
in the gravitational potential that one would infer from the observed mass distribution using GR (i.e. Newtonian gravity).
Thus,  in spiral galaxies one observes that rotation curves are flat functions of radius outside the region where
the luminous mass is concentrated, instead of falling as $r^{-1/2}$ as Newtonian gravity predicts.  Similarly,
clusters are observed to have X-ray gas that is far too hot to be bound in the Newtonian potential of the galaxies and
intracluster gas that comprise the visible cluster. 
Indeed this missing-gravity problem has been verified in far too many systems for it to be appropriate to review here.  
It is perhaps noteworthy that this same problem does not apply to denser systems of the same mass as
dwarf galaxies, such as globular clusters.

This missing gravity problem also rears its head in the question of how the large scale structures that we currently
observe in the universe formed.    In the early universe, radiation dominates the energy density;
however, the energy density of radiation $\rho_{r}\propto(1+z)^4$ and so matter eventually comes to dominate, even
though the number density of photons is approximately $10^{10}$ times that of baryons.   
During radiation domination, fluctuations in the energy density do not grow.
Loosely, in the competition between growth of fluctuations under the influence of their own self gravity,
and the damping of those fluctuations because of the expansion of the universe, the self-gravity loses.
Once matter dominates, fluctuations do grow, but  the density contrast increases only linearly as $(1+z)$.
However, we have a snapshot of the universe -- the CMB at redshift of $~10^3$ -- 
and it shows fluctuations of only $~10^{-5}$.
It is thus impossible for non-linear structures such as galaxies and clusters
to have grown in a SMGRC until today, much less by redshifts of a few when
they are observed. Put another way -- there is insufficient attractive self-gravity in SMGRC to explain
the growth of the observed structures in the universe. 

The above failures of SMGRC are failures to provide sufficient ordinary attractive gravity, but attractive
gravity alone is insufficient to explain the universe we see.  In particular both the universe's current expansion
and its  expansion at some epoch in the very distant past  seem to involve acceleration.    The current
accelerating expansion is inferred by direct observation of the recession velocity (redshift) of distant
type 1a supernovae as a function of their distance, as derived from their apparent brightness.  
(Originally Riess 1998 and Perlmutter 1999. Most recently 
Amanullah 2010,  Kowalski 2008,  Kessler 2009, Contreras 2009,
Balland 2009, Bailey 2008 and Hicken 2009)
(These supernovae are standard candles once one recalibrates  to account  for their different masses.)   

This accelerated expansion is also necessary to explain the age of 
our universe whose current hubble expansion rate is $H_0\simeq 72{\rm km/s/MpC}$.  
A flat matter dominated universe of this expansion rate is less than 10 billion years old,
whereas the oldest globular clusters are thought to be 2-3 billions years older than that.
An accelerating universe would be older than the objects in it.

An earlier period of accelerated expansion is thought to be necessary to explain
other conundrums about the universe in which we live.   Most prominent among these is
the homogeneity problem: if light of the CMB is just reaching here from after traveling
through the universe since it was a brief fraction of its current age, then how is it that
the CMB is a blackbody of temperature 2.7K in all directions?  Shouldn't the emitting
regions on opposite sides of the sky (say) be causally disconnected, and so at unrelated
temperatures.  This and a number of similar conundrums -- the geometric flatness 
of the universe, its age, the high entropy -- can all be addressed if the universe underwent
a period of accelerated expansion, called inflation,  early in its history.

The attractive gravity among
particles of matter or radiation acts to deccelerate the expansion of the universe. 
 Within general relativity,
acceleration is driven by forms of energy density that scales at least as slowly as
$(1+z)$.   The Standard Model has only one such form of energy density -- the energy of the vacuum, $\rho_v$.
It is unclear within the Standard Model what is the value of $\rho_v$.  A naive estimate
is approximately  $10^{120}$ times what is required to drive the current
cosmic acceleration.    This is known as the cosmological constant problem.  
Certainly the model does not contain energy density to drive both the early universe
inflation, and the current acceleration.

Finally, we mention briefly one other possible failure of SMGRC: the anomalous
decceleration of the Pioneer 10 and 11 spacecraft.
This anomalous acceleration of $a_P = (8.74 \pm 1.33) \times 10^{-10} {\rm m/s}^2$
appeared in the orbit of the  spacecraft once they entered the outer solar system
and persisted until contact with them was left.
It has long been claimed that no known force can provide the decceleration.  
Recently, a prosaic explanation has been suggested for at least part of the anomalous acceleration (Turyshev 2010), 
but that remains preliminary.   

\section{The Standard Solution -- Three Dark Sectors}

The standard solution to this serial failure of the SMGRC is well known to the reader,
and is known as the LambdaCDM ($\Lambda$CDM) concordance inflationary cosmology.
It involves introducing three entirely separate dark sectors:   dark matter (non-relativistic
particles),  dark energy, and an inflaton (a second form of dark energy).     

The inflaton is a scalar field introduced to address the homogeneity and flatness problems through
accelerated expansion in the early universe.  This accelerated expansion is the result
of a period where the potential energy of the inflaton field is the dominant source of energy density
and pressure in the universe.   Because the pressure is negative, or equivalently because
potential energy density does not dilute as the universe expands if the field is not evolving
(no kinetic energy), the Friedmann equation \ref{eqn:Friedmann}
implies exponential expansion.      The great success of the inflationary model is that even
as it explains the homogeneity and flatness, it provides a mechanism for generating 
fluctuations that later seed the growth of large scale structure.   It is unfortunate that the
natural amplitude of those fluctuations is ${\cal O}(1)$, about five orders of magnitude larger
than is observed.   Taming those fluctuations is possible through mechanisms of varying
(technical) naturalness that fine tune the parameters of the inflaton action.   The disagreeable nature
of this fine tuning has, over time, largely been forgotten given the success of the model in 
reproducing the Harrison-Zeldovich (scale free) spectrum of adiabatic perturbations that
had previously been hypothesized.   The dominance of adiabatic fluctuations
on a wide range of scales is consistent with inflation, though not a necessary feature of
inflationary models.  The phase coherence of those adiabatic modes evidenced by the 
observed doppler peaks in the CMB is characteristic of inflation.
The moderately generic inflationary prediction that
the spectrum deviates somewhat from scale free either has been confirmed, or may soon
be tested; similarly the generically predicted  tensor perturbations are being sought, though
their amplitude may be too small to be detectable.     

Efforts to connect the inflaton to  Standard Model, or "just beyond the Standard Model" physics 
(eg. the Minimal Supersymmetric Standard Model) have not been successful, so that the
range of candidate inflatons is enormous but with little or no foreseeable connection to physics that
can be tested in the laboratory.

The standard cosmology requires potential energy to dominate not just in the early universe, but
again today. Today's dark energy density's properties are little better determined than those of the inflaton -- 
it must be approximately three-quarters of the  critical density or about $(10^{-3} {\rm eV})^4$ --
many many many orders of magnitude smaller than the inflaton energy density  (except, see Dutta 2005).
In the simplest model, it is just a cosmological constant, aka the vacuum energy density, with isotropic pressure,
an equation of state $w\equiv p/\rho = -1$ and no anisotropic stress.   
All observations to date are consistent with this simplest model, although
that does not deter theorists from exploring "richer" (i.e. more complicated) models, nor
observers from seeking to constrain those models.

Finally (though historically first in priority), standard cosmology requires the introduction of dark matter,
in particular non-relativistic (a.k.a. cold) dark matter.  Numerous candidates have been identified.  They
range in mass from fractions of an electron volt (eg. axions, which are cold because they are produced as
a bose condensate) to many solar masses.  Some are ad hoc (i.e. invented exclusively or at least principally
to solve the missing matter problem, with no connection to the standard model, nor to extensions of it
that were devised to explain particle physics problems.  Others are better motivated -- for example
the lightest supersymmetric partner (LSP) or the axion.   The first great success of dark matter is that it
addresses multiple problems simultaneously.   With sufficient dark matter appropriately distributed
the rotation curves of spiral galaxies, velocity dispersions of elliptical galaxies and kinematics (including X-ray
temperatures of clusters can be explained. The amount of dark matter that must be introduced is specific to
each system, as is its precise distribution, but the variation is not unexpected. Beyond that, the existence of
dark matter allows us to understand how large scale structure can grow by gravitational accretion even
though the observed fluctuations in the CMB are so small.  This is because dark matter fluctuations 
start growing from the equality of the matter and radiation energy densities -- 
earlier than baryonic matter fluctuations, which are limited by the coupling between matter and radiation
until recombination.  The dark matter fluctuations can therefore act as seeds for the growth of structure in the
baryonic matter.   

The dark matter model made at least two further successes in two predictions 
that have since been confirmed.
First, that light propagating past mass concentrations like clusters would be gravitationally lensed, and if one
used GR to extract the mass (distribution) of the cluster, it would be consistent with the mass distribution
as extracted by measurements of the kinematics of the constituent galaxies and gas (Tyson 1984).   Second,
that when structures dominated by dark matter (such as clusters) collided with each other, the dark
matter halos (and stars) associated with each would scatter (essentially) only gravitationally, even
as the gas associated with each had more complex interactions.  This prediction is consistent
with observations of systems like the Bullet Cluster (Clowe 2006).  

We will keep these three predictions  in mind as we explore the viability of alternatives to a multi-component dark sector.

\section{Modifying Gravity Instead}

The subtext of the previous section was that while our universe can apparently be modeled phenomenologically
by a multi-component dark sector, it cries out for a more parsimonious description.   Adding dark matter or dark
energy represents adding new forms of stress-energy $T_{\mu\nu}$, not included in the standard model, 
but preserving the Einstein field equations themselves -- the dynamical equations in which matter sources gravity:
\be
G_{\mu\nu} = 8\pi G_N T_{\mu\nu} .
\ee
However,  given that we have yet to find a model of the dark sector that elegantly wraps together dark matter, 
dark energy and inflaton, we might ask whether a modification of the dynamical equation itself (or more broadly
of the underlying theory of GR) could give us a model of the Universe that is shaved more closely by
Occam's Razor.   Of course it would be advantageous if it was not only a simpler model, but could be 
experimentally or observationally differentiated from  modifications of $T_{\mu\nu}$ exclusively.

There is at least some hint that such an effort at a unified theory of the deviations of the universe from SMGRC
might succeed. That is that there is a common scale appearing in several of these deviations. 
The observed cosmic acceleration clearly sets a scale -- $c/H_0\simeq4000$MpC. Alternately, this
can be viewed as an acceleration scale $H_0 c = 7\times 10^{-8}{\rm cm/s}^2$.
Interestingly, this is not so different from the typical value of the centripetal acceleration at which the 
rotation curves of galaxies deviate from the Newtonian (i.e. GR) prediction.  
Thus, the acceleration scale that MOdified Newtonian Dynamics (MOND), 
a phenomenological modified-gravity replacement for dark matter (Milgrom 1983), 
introduces is $a_0 \simeq 6 H_0c$.
Amusingly, the anomalous acceleration of the Pioneer satellites (Anderson 2002) was also of this order,
$a_{Pioneer} \simeq 9  \times 10^{?8} {\rm  cm/s}^2$.

One can even perhaps squint and see hints of this scale in the inflationary theory.  
The inflationary prediction that all fluctuations in the universe are (seeded by) realizations
of a Gaussian random statistically isotropic field, while consistent with observations 
at most angular scales, seem to fail at angular scales above $60^o$ (Spergel 2003,  Copi 2006).
In particular the two point-correlation function of the CMB is close to vanishing on all
angular scales between about $60$ and $170$ degrees. Because of cosmic variance,
this is very unlikely if the coefficients of the spherical harmonics, $a_{\ell m}$,
are Gaussian random.     Interestingly, $60^o$ is the angle subtended by the length
scale $H_0^{-1}$ at $z~1$.

The truth is that most modifications of gravity do not tie together all of these different failings of SMGRC,
and so they are, at least in that sense, no better or worse  than dark sector solutions.  It may well
be that instead of dark energy we have "accelerated expansion from gravity leaking to extra dimensions" 
(Deffayet 2001) as in DGP gravity (Dvali 2000), or a modest modification 
of the gravitational part of the Einstein-Hilbert action as in $f(R)$ models
(Carroll 2004 and references thereto).  

It is without a doubt a worthy goal to seek out  modified gravity models of dark energy or inflation
and look for points of observational differentiation from dark energy.  But if the aim is to seek
a model that provides a more comprehensive answers to the failures of the SMGRC,
capitalizing perhaps on the coincidences of scale in the observed anomalies, then
it will be necessary to address the replacement of dark matter by modified gravity.
A preliminary framework for considering this possibility is MOND (Milgrom 1983).  There is more
than one version of MOND, some modifying Newton's inverse square law of gravitation 
(or  Poisson's equation), some Newton's second law of motion (Milgrom 2005).   We shall frame our
discussion in terms of a modification of Poission's equation:
\begin{equation}\label{eqn:ModPoisson}
\nabla \cdot [\mu (  \abs{\vec{\del}\Phi}/a_0   )  \nabla \Phi ]=4\pi G \rho \,.
\end{equation}
\noindent
The function, $\mu(x)$ must be chosen to satisfy the necessary limits -- namely that $\mu(x)\to 1$ for large $x$ 
(the Newtonian limit) and $\mu(x)\to x$ as $x\to0$ (the MOND limit).  
Far from a static point mass, this MOND limit gives
$\nabla\Phi \to \sqrt{G M a_0}/r$.  For circular orbits, this means that in MOND
\be
\label{eqn:MONDorbitalv}
v_{\rm circular} \to \left(G M a_0\right)^{1/4}\,,
\ee
i.e. rotation curves are flat at large radii, as observed.
Surprisingly, the form
\begin{equation}
\label{eqn:standardmu}
\mu(x)=\frac{x}{\sqrt{1+x^2}} 
\end{equation}
works very well for most galaxies.   

What does work well mean?  First of all, \ref{eqn:ModPoisson} together with \ref{eqn:standardmu} 
reproduces the broad brush features of  galactic rotation curves.  
Of course this is not surprising -- it was chosen to give Newtonian gravity in the inner reaches
where there are a lot of baryons,  and flat rotation curves outside the baryons.  But there is a little bit
more to even this simple statement.  After all, this this simple phenomenological model has one
universal function $\mu(x)$ and only one adjustable constant per galaxy -- the mass-to-light ratio.
This is considerably less freedom than is available in fitting dark matter models to individual 
galaxies, as there one has an unknown dark matter density distribution (limited principally
by expectations that such distributions be both reasonable (eg. non-singular) and 
reasonably similar among galaxies of similar morphology,
and hopes that they could emerge from numerical modeling of galaxy formation.    
Moreover, the mass-to-light ratios that are derived for spiral galaxies from MOND are
eminently reasonable -- in other words they are around $20$, consistent with expectations
independent of MOND.

Furthermore, MOND gives us a way to understand the well known Tully Fisher relation, that
the luminosity of a spiral galaxy scales as the fourth power of its velocity width -- the amplitude
of its rotation velocity.  After all \ref{eqn:MONDorbitalv} implies that $M\propto v^4$, 
so if mass-to-light ratios of spirals are reasonably uniform, then ${\cal L}\propto v^4$ as well.   
Similar arguments (esp. Sanders 2010) have been made about the Faber-Jackson relation -- the empirical relation 
between the luminosity of an elliptical galaxy and its velocity dispersion:  ${\cal L}\propto \sigma^4$.

Even more surprisingly the agreement of MOND with galaxy rotation curves 
is not only at the qualitative level.  Galaxy rotation curves have all sorts of bumps and wiggles.
In a dark matter model, such features can always be attributed to features in the dark matter
distribution.   Not so in MOND -- all features must be traced back to the baryon distribution, which
alone sources the gravitational field.   Amazingly (at least to the author), MOND succeeds
in correctly reproducing the details of many (or perhaps all?) rotation curves from the details 
of the distribution of baryonic mass (Sanders and McGaugh 2002; McGaugh 2005).

The situation for MOND at smaller scales  is less clearly positive
than at intermediate scales.  Although dwarf galaxies
and globular clusters are of similar mass, dwarf galaxies have a missing gravity
problem, and globular clusters, it has been argued, do not (Baumgardt  2009; Jordi 
2009; Lane 2010; Conroy 2010).   Of course, the mass of the system is
not the only issue, especially when the system is not isolated, such as for most globular
clusters embedded in the Galactic halo.  The concentration of the system also matters,
as does the environment -- especially at radii where the Newtonian acceleration falls
below the MOND threshold.
Thus others (Gentile 2010) find that such globular clusters are fully consistent with MOND,
and may even be demonstrably inconsistent with Newtonian gravity (McGaugh 2010).

At larger scales, it has long been known that (galaxy) clusters do not follow MOND.   
For example Aguirre {\it et al.} (2001) argued that 
``given observed gas density and enclosed mass profiles and the assumption of hydrostatic equilibrium, 
MOND predicts radial temperature profiles which disagree badly with observations.''
They showed  this explicitly for the Virgo, Abell 2199 and Coma clusters, but argued that
the results are general.  The discrepancy could be resolved by the presence of 
additional non-luminous baryons, but it would have to be 1-3 times the observed cluster gas mass 
If this discrepancy is to be resolved by positing additional (presumably baryonic) dark matter, 
then this dark matter must at least equal the observed mass within 1 Mpc.     Of course, one
could just posit the presence of dark matter in the cluster.   This may seem like a cop-out -- the
whole point of MOND is to {\em avoid} introducing new unknown forms of stress energy -- 
but is not quite as bad as the first impression would suggest, because it could be hot  dark matter.
(Dark matter that was moving sufficiently fast at early times not to be captured into galaxies.)
Why is this less distateful than cold dark matter?  Because at least there are natural candidate --
neutrinos.  

Although neutrinos are massless in the standard model, we know that neutrinos are not properly
described by the standard model.  
In particular the eigenstates of flavour (i.e. electron, mu and tau neutrinos) are not the freely propagating 
eigenstates of the Hamiltonian.   The canonical models for this neutrino mixing require that
the neutrinos have non-vanishing masses.  It had been suggested (Angus 2007) that a neutrino mass of $2-7$eV would 
appropriately fill the need of clusters for dark matter in addition to MOND,  however (Natarajan and Zhao)
have since argued against that model's sufficiency to explain observed weak lensing signals from clusters.
Nevertheless, it continues to be argued (Feix) that, at least within modified gravity theories with appropriate MOND
limits, neutrino mass can play the role demanded.

Finally, before we proceed to discuss some serious concerns, it should be noted that
since the  Bekenstein's introduction of TeVeS (Bekenstein 2004), there exist covariant
theories based on an action principle that recover MOND (in its modified Poisson equation form) 
in the appropriate limits. TeVeS is a theory with.

In Bekenstein's theory gravity is mediated  by a tensor field $\tilde{g}_{ab}$ with associated 
 connection $\tilde{\nabla}_{ab}$ and inverse $\tilde{g}^{ab}$, a one-form field $A_{a}$, 
 and a scalar field $\phi$. The Einstein-Hilbert action governs the dynamics of  $\tilde{g}_{ab}$: 
\begin{eqnarray}
S_{\tilde{g}}= \frac{1}{16\pi G}\int d^{4}
x(-\tilde{g})^{\frac{1}{2}}\tilde{R} \nonumber
\end{eqnarray} 
(G is Newton's constant and $\tilde{R}$ is the scalar curvature 
of $\tilde{g}_{ab}$.) The dynamics of $\phi$ are given by
\begin{eqnarray}
S_{s} = -\frac{1}{16\pi G}\int d^{4}x(-\tilde{g})^{\frac{1}{2}}
\left[\mu(\tilde{g}^{ab}-A^{a}A^{b})\tilde{\nabla}_{a}
\phi\tilde{\nabla}_{b}\phi+V(\mu)\right] \nonumber
\end{eqnarray}
where we have used the convention of Skordis (2006).
$\mu$ is a non-dynamical field and V is a free function that is chosen 
to give TeVeS a non-relativistic MONDian limit. 
The dynamics of $A_{a}$ are described by
\begin{eqnarray}
S_{v} = -\frac{1}{32\pi G}\int d^{4}
x(-\tilde{g})^{\frac{1}{2}}\left[KF^{ab}F_{ab}
-2\lambda(\tilde{g}^{ab}A_{a}A_{b}+1)\right] \nonumber
\end{eqnarray}
where $F_{ab}= 2\tilde{\nabla}_{[a}A_{b]}$ and brackets denote 
antisymmetrization. Indices are raised with $\tilde{g}_{ab}$ and 
K is a dimensionless parameter. Variation with respect to 
the Lagrange multiplier field $\lambda$ enforces that $A_a$ be `unit timelike':
$\tilde{g}^{ab}A_{a}A_{b}=-1 $.

A key feature of TeVeS, aside from the breaking of Lorentz symmetry by the
background time-like $A_a$, is that matter couples to a different metric $g_{ab}$ 
than the dynamical one $\tilde{g}^{ab}$ described above. Test particles
thus follow the geodesics of $g$ not $\tilde{g}$.
For some collection of matter fields $f^{A}$ the matter action is thus
\begin{eqnarray}
S_{m} = \int d^{4}x(-g)^{\frac{1}{2}}L\left[g,f^{A},
\partial f^{A}\right] \nonumber
\end{eqnarray}

When the unit-timelike constraint on $A_{a}$ is satisfied,
the metrics, one-form field, and scalar field are related by the disformal transformation:
\begin{equation}
\label{md}
\tilde{g}_{ab} = e^{2\phi}\left(g_{ab}+2A_{a}A_{b}\sinh(2\phi)\right) ..
\end{equation}

Zlosnik {\it et al.} showed (Zlosnik 2006) that TeVeS can be re-written as a Vector-Tensor theory 
akin to Einstein-Aether theories (see for example Eling 2004 for a review) 
with non-canonical kinetic terms.  
This enabled them (Zlosnik 2007)
to broaden the class of covariant modified gravity 
models that have similar properties to TeVeS --  appropriate non-relativistic MOND (and Newtonian limits).
They all share the existence of a timelike vector field $A_a$, a so-called Einstein Aether, and hence
are referred to as Generalized Einstein Aether (GEA) theories.  

A subclass of these models are the so-called Generalized Einstein Aether 
 GEA theories (Zlosnik 2007) that start with a covariant
action and recover MOND (in its modified Poisson equation form) in the appropriate limits.
In its simplest form,  GEA contains  a vector field, \textbf{A} coupled to Einstein-Hilbert gravity  but not to matter:
\begin{eqnarray}
S=\int d^4x \sqrt{-g}\left[\frac{R}{16\pi G_N}+{\cal L}(g,A)\right]
+S_{M} 
\end{eqnarray}
where $g_{\alpha\beta}$ is the metric, $R$ the Ricci scalar of that metric.
$\cal{L}$ is constructed to by generally covariant and local. 
For simplicity, the matter action, $S_M$, is taken to couple to \textbf{g} but {\it not}
to \textbf{A}. 

As for TeVeS, \textbf{A} is enforced to be unit time-like through a Langrange multiplier.
Otherwise, the Lagrangian is taken to depend only on covariant derivatives of $A$,
\begin{eqnarray}
\label{eq:Lagrangian}
{\cal L}(A,g)&=&\frac{M^2}{16\pi G_N}
	 {\cal F}({\cal K}) +\lambda(A^\alpha A_\alpha+1) 
	\nonumber \\
{\cal K}&=&M^{-2}{\cal K}^{\alpha\beta}_{\phantom{\alpha\beta}\gamma\sigma}
\nabla_\alpha A^{\gamma}\nabla_\beta A^{\sigma} \nonumber \\
{\cal K}^{\alpha\beta}_{\phantom{\alpha\beta}\gamma\delta}&=&c_1g^{\alpha\beta}g_{\gamma\sigma}
+c_2\delta^\alpha_\gamma\delta^\beta_\sigma+
c_3\delta^\alpha_\sigma\delta^\beta_\gamma
\end{eqnarray} 	
where $c_i$ are dimensionless constants and $M$ 
has the dimension of mass. 
(TeVeS is formally equivalent to a  theory with an extended ${\cal K}$,
and  a more exotic method of achieving a
 non-vanishing background value for \textbf{A}.)
 More details can be found in (Zlosnik 2007).

TeVeS and GEA possess important advantages over MOND alone.  Most importantly they are (at least nominally)
comprehensive frameworks in which any question about classical gravity can in principle be answered.  Thus
it is appropriate to ask questions about cosmology, stellar structure, compact objects, gravity waves, gravitational
lensing, solar system dynamics.   The answers may be difficult to obtain in practice, but unlike in MOND one cannot
respond to disagreements with data by invoking the fact that  the model  strictly applies only to isolated/static/spherically 
symmetric systems.   

 It is encouraging  that within this covariant framework 
 one can simultaneously realize both the non-relativistic MOND limit 
 and a cosmology in which the scale factor evolution is consistent with observations 
 -- i.e. late-time acceleration, and  inflation.
(For GEA, see Zlosnik 2008; for TeVeS see Skordis 2005 and Diaz-Rivera 2006.)
One of the reasons that it is indeed relatively easy to accomodate both MOND
on galactic scales and acceleration on cosmological scales is that they
tend to probe different domains of the arbitrary functions that characterize the theory.
For example, in GEA, for a universe described by the usual flat FRW metric, filled with a perfect fluid,
\be
\label{KappaofH}
{\cal K} = 3\frac{\alpha H^{2}}{M^{2}} 
\ee
where, of course, $H \equiv \frac{\dot{a}}{a}$,
the dot denotes differentiation with respect to $t$, and,
following  $\alpha \equiv c_{1}+3c_{2}+c_{3} < 0$. 
$\alpha$ is negative and hence so is ${\cal K}$.

The modified $00$ Einstein equation is (Zlosnick 2008)
\be
\label{00m}
[1-{\cal F}_{{\cal K}}\alpha]H^{2}+\frac{1}{6}{\cal F}M^{2} = \frac{8\pi G}{3}\rho
\ee
where ${\cal F}_{{\cal K}} \equiv d{\cal F}/d{{\cal K}}$.
Restricting  to ${\cal F}$ of the form
\begin{eqnarray}
\label{eqn:FofK}
{\cal F}=\gamma (-{\cal K})^n .
\end{eqnarray}
 equation \ref{00m} is more clearly a modified  Friedmann equation:
\begin{eqnarray}
\label{modifiedFriedmann}
\left[1+\epsilon\left(\frac{H}{M}\right)^{2(n-1)}\right]H^2=\frac{8\pi G}{3}\rho  ,
\end{eqnarray}
where 
\be
\epsilon=(1-2n)\gamma(-3\alpha)^{n}/6. \nonumber
\ee

We see that, for $n=1/2$  the Friedmann equations are unchanged ($\epsilon=0$).  
For $n=1$  $\epsilon=\gamma\alpha/2$ and the only effect (on scale factor evolution) is that
Newton's constant is renormalized  -- $G'=G/(1+\epsilon)$.
For $n=0$, we recover a cosmological constant, $\Lambda\simeq sign(-\gamma)  M^2$.
More generally (Zlosnick 2008), there is late time accelerated expansion whenever $\gamma(2n-1)>0$.
Moreover, not only can one obtain late time acceleration, but if $n$ evolves as $K$ evolves,
then one could obtain a wide range of scale factor evolutions. 

A particular key point however is that cosmology probed only the ${\cal K}<0$ behaviour of ${\cal F}$.
This is because it is sensitive to the time derivatives of $A$ not its spatial derivatives.  However, in
the MOND limit, for  example in a galaxy, the system is  static. 
Thus it is is the spatial gradients of $A_a$ that dominate ${\cal K}$,  and thus ${\cal K}>0$.

It is in some ways unfortunate that structures and scale factor evolution probe disjoint ranges of ${\cal K}$.
It means that it is difficult to promote such theories as offering truly unified alternatives to dark matter
plus dark energy.  Rather one should say that both can be accommodated, and the fact that the relevant
acceleration scale associated with  each  is roughly similar just means that it is easier to interpolate between  
the two domains of ${\cal K}$.

\section{Troubles with Modifying Gravity}

Having promoted the virtues of modifying gravity, it is time to get to the point of this talk  -- that it is hard to modify gravity
and get away with it.  We shall focus on a series of challenges that any attempt to 
replace the dark sector by modified gravity must face.

\subsection{When Modified Gravity Means Dark Matter}

One of the successes we outlined for dark matter is that it allows large scale structures to grow
in the early universe, and without the dark matter,  there is insufficient time to allow structure to grow.
This is of course obvious, but there has always been the vaguely justified hope that, since MONDacts
to strengthen gravity on large scales the increased attractiveness would result in more rapid structure
growth.     This of course ignores the role that stronger gravity plays in damping structure growth -- faster
growth necessitates that the attractive gravity work still harder to get structure to grow rather than be damped
out by faster Hubble growth.   Proving there is a no go theorem is not possible in the context of MOND because
MOND is not well defined on cosmological scales.     Nevertheless Lue and Starkman (Lue 2004b)
were able to make some   clear progress on this front relatively generically.   
As they write: \newline\indent
\vbox{``By insisting that the gravitational interaction that accounts for the
Newtonian force also drives cosmic expansion, one may kinematically identify
which cosmologies are compatible with MOND, without explicit reference to
the underlying theory so long as the theory obeys BirkhoffÕs Theorem. Using this
technique, we are able to self-consistently compute a number of quantities
of cosmological interest. We find that the critical acceleration a0 must have
a slight source-mass dependence ($a0 \tilde M^{1/3}$) and that MOND cosmologies
are naturally compatible with observed late-time expansion history and the
contemporary cosmic acceleration. However, cosmologies that can produce
enough density perturbations to account for structure formation are contrived
and fine-tuned. Even then, they may be marginally ruled out by evidence of
early ($z \lsim 20$) reionization}

The calculation is reasonably straightforward.   It generalizes the classic elementary derivation 
of the Friedmann equation from conservation of energy.    We reproduce its outlines here. (For details see Lue 2004a.)

Consider a homogeneous universe filled with homogeneous matter of density  $\rho(t) \sim a^{-3}$.
Assume that it is  described by the (flat FRW) line element
\be
        ds^2 = dt^2 - a^2(t)\delta_{ij}dx^idx^j\ ,
\label{line-cosmo}
\ee
with some specified scale-factor evolution, $a(t)$.
Gravitational dynamics must be altered in a specific way to
achieve the observed cosmic expansion history, $a(t)$.  
Writing   $a(t)$ as the solution to some alternative Friedmann equation:
\be
        \dot{a}^2/a^2 = H_0^2g(x)\ ,
\label{FRW}
\ee
(where $x = {8\over 3}\pi G\rho/H_0^2$ is a dimensionless density
$G$ is Newton's constant, and $H_0$ is today's Hubble scale), we see
that the function $g(x)$ is determined by the observed $a(t)$.  
If one requires that the fundamental gravitational theory respects Birkhoff's Theorem, 
then one can uniquely determine the metric of a spherically symmetric
source (Lue 2004a).  That metric is described by the line element
\be
        ds^2 = g_{00}(r)dt^2 - g_{rr}(r)dr^2 - r^2d\Omega\ , 
\ee
\be
        {\rm with}\ \ g_{00}(r) = g^{-1}_{rr} = 1 - r^2H_0^2g\left(r_g/r^3H_0^2\right)\ .
\label{metric}
\ee
Here $r_g = 2GM$, is the usual Schwarzschild radius of a matter
source of mass $M$.  Note that the metric components
are completely determined by $a(t)$.

In MOND the gravitational acceleration exerted by a body of mass $M$ obeys 
\be
        a = 
        \begin{cases} 
        -{1\over 2}{dg_{00}\over dr} = -GMr^{-2}  & \vert a\vert > a_0 \cr
        \sim  -r^{-1} & \vert a\vert < a_0 \cr 
        \end{cases} \label{preMOND}
\ee
for some critical acceleration, $a_0$.  
To recover  equation (\ref{preMOND}), we must have
\be
        g(x) = \begin{cases}
                x + c_1x^{2/3} &  {\rm Einstein} (x>x_c)\cr
                \beta x^{2/3}\ln x + c_2x^{2/3} & {\rm MOND} (x<x_c) , \cr
        \end{cases}
\label{g}
\ee
for some constant parameters, $\beta$, $c_1$, and $c_2$.    Thus
\be
a = \begin{cases}
-{1\over 2}{r_g\over  r^2} & \vert a\vert > a_0 \cr 
-{3\beta\over 2}{(r_gH_0)^{2/3}\over r} & \vert a\vert < a_0\ , \cr
\end{cases}
 \label{aMOND} \ee
where the critical MOND acceleration, $a_0$, is
\be
        a_0 = H_0\left[9\beta^2(r_gH_0)^{1/3}\right]\ .
\label{a0}
\ee
Observationally, we choose $\beta \approx 15$ so that for source
masses the size of large galaxies ($M\sim 10^{11}M_\odot$), the
critical acceleration is $a_0 \approx {1\over 6}H_0$.
To ensure that $g(x)$ is continuous at $a = a_0$,
$c_1 = c_2 + 3\beta\left[\ln (3\beta)-1\right]$.
The remaining constant represents an arbitrary choice in zero-point
energy for the Newtonian potential.
  Although $c_1$ and $c_2$ do not affect the Newtonian
  acceleration, they simulate curvature-type terms in the Friedmann equation 
despite the spatially-flat cosmology (see Eq.~(\ref{line-cosmo})).  
However, the resulting change in $g_{rr}$ (see Eq.~(\ref{metric}))  causes 
only immeasurably small effects on gravitational lensing and
  the post-Newtonian parameter, $\gamma$.

The form Eq.~(\ref{aMOND}) for the MOND gravitational acceleration is
slightly different than that typically considered, where
$a_0 = H_0/6$ is a universal constant.  The combination of 
homogeneous cosmology and Birkhoff's Theorem compels 
a weak dependence of $a_0$ on the source mass, $a_0\sim M^{1/3}$.
The confrontation of this prediction with data will be explored in (Kafka 2011).
More immediately, this combination
determines both the modified Friedmann equation, Eq.~(\ref{FRW}), and
the full Schwarzschild-like metric of a spherical mass source,
Eq.~(\ref{metric}), without explicit reference to the details of the
fundamental theory.  With these two governing relationships, one may
 compute modifications of planetary ephemeris, gravitational lensing, growth of density perturbations , 
 and determine the late-time integrated Sachs--Wolfe (ISW) effect on the CMB.  
 Details of these calculations for arbitrary $g(x)$ are given in (Lue 2004).

Here we recall particularly on accommodating late-time acceleration into
MOND, and on the growth of fluctuations.  Indeed, to obtain late time
acceleration we must modify the form of $g(x)$ at very low $x$:
\be
        g(x) = \begin{cases}
         x + 3\beta x^{2/3}\left[\ln(3\beta)-1\right] & x \gsim (3\beta)^3  \quad\quad\quad\ \ (Einstein)
, \cr
         \beta x^{2/3}\ln(1+x) &   0.1 \lsim x \lsim (3\beta)^3 \quad  (MOND)
, \cr
         \Omega_\Lambda & x \lsim 0.1  \quad\quad \quad\quad\ \ (``Dark Energy'')
, \cr
        \end{cases}
\label{g-full}  
\ee
where $g(x)=\Omega_\Lambda \approx 0.7$ is equivalent to a cosmological constant.
Interestingly, as discussed in (Lue and Starkman 2004),  
if $\beta$ were an order-of-magnitude larger or smaller, 
one could not extend the MOND regime all the way to the deceleration-acceleration transition and
still be able to maintain both $H \sim H_0$ as well as $\ddot{a}/a\sim H_0^2$.
On the other hand, this is a rather complicated form for $g(x)$, and there
is no particular reason to expect this to emerge naturally from a covariantized MOND theory.

Equations~(\ref{FRW}) and~(\ref{g-full}) allow us to compute the growth
of linear perturbations in this cosmology.  In the absence of dark matter, 
we must wonder whether MOND has sufficient growth from recombination 
to produced the observed structures at $z\simeq few$.
As discussed in (Lue and Starkman 2004), the evolution of linear
density perturbations can be obtained in closed form.
Consider a uniform overdensity in a localized spherical region such that
\be
        \rho(t) = \bar{\rho}(t)\left[1+\delta(t)\right]\ ,
\ee
where $\bar{\rho}$ is the background matter density that
follows cosmological evolution.  Parameterizing time-evolution
using $x = 8\pi G\bar{\rho}/3H_0^2$, the growing
perturbation mode $\delta(x)$ goes as
(Lue 2004 and Multamaki 2003):
\be
     \delta(x) \propto  g^{1/2}(x)\int_x^\infty {dy\over y^{1/3}g^{3/2}(y)}\ .
\label{delta}
\ee
The normalization is fixed by requiring that   $\delta \sim few \times 10^{-5}$ at recombination.
In pure matter-domination for Einstein FRW, $\delta \propto x^{-1/3} \propto a(t)$.  
Even if growth were as large as this the growth of density perturbations would be
insufficient to account for the observed structure formation.  
But in  each of the three regimes in Eq.~(\ref{g-full}), growth is slower than that given
benchmark, $\delta < a(t)$.  
It may seem counterintuitive that growth is suppressed during the MOND regime, 
when the self-gravitation of overdensities is enhanced, but 
the stronger gravity drives faster expansion. 
This suppresses perturbation growth more than the stronger self-gravity enhances it.

This  suppression of structure growth would seem to pose a significant difficulty for
cosmological incarnations of MOND.   One might point to the fact that in "Real MOND"
$a_0$ does not scale with $M$, and so hope that these problems are peculiar to
the above calculation.  However,  since that relied only on the validity of Birkhoff's Theorem,
which we shall argue below one wants to relinquish as gently as possible if at all,  it
seems unlikely that this conclusion can be avoided without significant consequences.

In light of the previous argument, it was surprising (at least to the author) to learn
that in TeVeS and GEA the growth of structure is not suppressed.  However,
 the very way that covariant versions of MOND evade this constraint is enlightening
and arguably undermines the MOND program.
Skordis {\it et al.} (2003) were the first to study the growth of perturbations in TeVeS. 
Their numerical analysis identified growing modes in the TeVeS scalar field,
which support the growth of fluctuations in the Newtonian potential.
Dodelson and Liguori (2006) showed analytically that in fact it was 
growing modes in the vector fields that were responsible for the growth
of structure in TeVeS.    Zlosnik {\it et al.} (2004) showed the same thing for GEA --
a vector instability could replace dark matter as a seed for the growth of baryonic
structure after recombination.   

We thus seem driven to the conclusion that MOND -- which was designed to 
avoid the addition of a new undiscovered degree of freedom that source
the gravitational field in galaxies and clusters today and in the linear perturbations
that are their precursors,, 
requires instead the introduction of a new undiscovered degree
of freedom that source the gravitational field in those  linear-perturbations.
This is in addition to the unexpectedly large neutrino masses that must be introduced
to explain the dynamics of clusters.
In the light of results of Lue {\it et al.} (2003 and 2004), it is difficult to see how 
this problem can be made to away.

Much has been made of the failure of MOND to model the Bullet Cluster
(as well as other non-static, non spherically symmetric systems), 
which has been taken as "... direct empirical proof of the existence of dark matter" 
(Clowe 2006 ).    Specifically, in these systems the gravitational weak lensing
centers are coincident with the stars, but not with the hot x-ray gas.  This is what
one would expect from a dark matter model.  If two clusters collide, the dark
matter halos, and the stellar distributions, should pass through each other, perturbing
each others' paths only slightly, through gravity.      Meantime, the hot  x-ray gas from the
two colliding clusters appear to have interacted non-gravitationally, and therefore slowed
down relative to the stars.   

More accurately, such systems might be portrayed as
confirming the predictions of the cold dark matter theory.  However, they should not
have been unexpected in the context of MOND.   After all,as we described, clusters
are known, even in MOND, to require some additional source of gravity both today and 
at the time of their formation.  Why then would we  expect the weak lensing signal to
coincide with the mass distribution?   Thus it is probably reasonable that one can find 
ways to recover the observed weak lensing signal of the bullet cluster in the
context of modified gravity theories (Dai 2008; Brownstein 2007).   As Dai writes regarding GEA
"As vector-field fluctuations are responsible in GEA for seeding baryonic structure formation, vector-field concentrations can exist independently of baryonic matter. Such concentrations would not be expected to be tied to baryonic matter except gravitationally, and so, like dark matter halos, would become separated from baryonic matter in interacting systems such as the bullet cluster. These vector-field concentrations cause metric deviations that affect weak lensing. Therefore, the distribution of weak lensing deviates from that which would be inferred from the luminous mass distribution, in a way that numerical calculations demonstrate can be consistent with observations. This suggests that MOND-inspired theories can reproduce weak lensing observations, but makes clear the price: the existence of a coherent large-scale fluctuation of a field(s) weakly tied to the baryonic matter, not completely dissimilar to a dark matter halo."
On the other hand, it is equally plausible that a specific modified gravity theory would require
the addition of cluster dark matter, as has been argued for TeVeS (Feix 2007).

But whether the missing gravity of clusters is caused by hot dark matter, or some dark field halo,
it is clear that  theories with MOND limits need to have physics beyond MOND -- dark degrees of freedom --
to explain clusters.

\subsection{Keeping cosmology out of your galaxy}

One difficulty for covariant versions of MOND that has not yet been well explored is the
tension between the MOND limit and cosmology (but see Jacobs 2011)
Recall from above that the (or at least a possible) starting point of MOND is to write a modified Poisson equation:
\begin{equation}\label{Mod Poisson}
\nabla \cdot \[\mu\(  \abs{\vec{\del}\Phi}/a_0  \)  \nabla \Phi\]=4\pi G \rho  .
\end{equation}
At least this is meant to be the correct non-relativistic static spherically symmetric limit of the correct
underlying covariant theory.    It is meant to apply to galaxies and clusters.   But cosmic structures
exist within a cosmic background which is decidedly not static.
Thinking of MOND naively in terms of \eqref{Mod Poisson}, one is inclined to ignore cosmology, as the ``Hubble acceleration" is roughly $H_0^2 r$, where $H_0$ is the Hubble constant today and $r$ is the distance to the center of the gravitating object (e.g. galaxy or cluster).  The same cannot be said, in general, for a corresponding covariant theory. In GEA
for example (the same is also true of  BIMOND) invariant scalars are built out of derivatives on the metric(s) and, in particular, this means that time derivatives are put on the same footing as gradients.  Due to cosmological considerations, therefore, one will need to compare the natural value of time derivatives ($\sim H_0$) with spatial derivatives of gravitational potential(s).  Since the MOND acceleration scale $a_0$ is of order $H_0$, serious tension is possible.

For example, consider two different metrics to approximate that of a galaxy,  
one static:
\begin{align}
g_{00}&=-(1 + 2\Phi)\label{h00_static}\\
g_{ij}&=(1-2\Psi) \de_{ij}\label{hij_static}
\end{align}
and the other dynamical (accounting for cosmological expansion):
\begin{align}
g_{00}&=-(1 + 2\Phi)\label{h00_dyn}\\
g_{ij}&=(1-2\Psi)a(t)^2 \de_{ij}\label{hij_dyn}
\end{align}
where, in both metrics, we assume $g_{0i}=0$.  
If we choose the static metric, \eqref{h00_static} and \eqref{hij_static}, 
we find the Poisson equation in GEA to be modified as
\begin{equation}\label{staticPoisson}
\vec{\del} \cdot \[ \(  2 + c_1 {\cal{F}}'\) \vec{\del} \Phi\] =8 \p G \rho
\end{equation}
with the scalar ${\cal{K}}$ given by
\begin{equation}\label{staticK}
{\cal{K}}=\frac{-c_1 (\vec{\del} \Phi)^2}{M^2} +{\cal{O}}(\Phi^3)
\end{equation}
Thus, in order to recover \eqref{Mod Poisson} 
in the limit $\vec{\del} \Phi \ll M$, one finds the free function to take the form
\begin{equation}\label{staticF}
{\cal F(K)} \sim \a {\cal K} + \b {\cal K}^{3/2}
\end{equation}
for some constants $\a$ and $\b$.  

Considering we do live in an expanding universe, however, it is arguably more correct to use a perturbed FRW metric.  While our experience tells us to expect that cosmology should be irrelevant when investigating e.g. solar system dynamics, at large separations it is clearly present, therefore nothing should stop us from using a metric which accommodates both limits.  
However, using the dynamic metric, \eqref{h00_dyn} and \eqref{hij_dyn}, one finds (Jacobs 2011) 
that  the modified Poisson equation is further modified:
\begin{align}\label{cosmoPoisson1}
M^2 {\cal F}\( \f{1}{2} + \Phi\) &+  3H^2\(1 - \a {\cal F_K}\)\notag\\
& + \vec{\del} \cdot \[  \(2 + c_1 {\cal F_K}\) \vec{\del} \Phi \] = 8 \p G\(\bar{ \r} + \de\r\) .
\end{align}
Also the scalar ${\cal{K}}$ is now given by
\begin{equation}\label{cosmoK1}
{\cal K}= \f{1}{M^2} \[ 3\a H^2\(1- 2\Phi + \Phi^2\) - c_1 \(\f{\vec{\del} \Phi}{a}\)^2\] .
\end{equation}
But in the regime of greatest interest $\vec{\del} \Phi$ is of order $H=H_0$. 
Thus, it appears that Hubble expansion is relevant at the very point where the MOND regime begins.
 It is not clear, and in fact seems unlikely, that the ${\cal F(K)}$ given by \eqref{staticF} would still be 
 sufficient for a successful theory of MOND.  
 
 We now see what the unexpected price of ``covariant-izing" MOND can be -- the same mechanism that allows one to modify the Poisson equation, namely using tensors built from derivatives of the metric, puts Hubble expansion on the same footing as gradients of the Newtonian potential(s).  Factors of $H_0$ that are introduced are non-negligible since MOND is necessarily concerned with the limit where $\vec{\del} \Phi$ is of order $H_0$. Cosmology apparently becomes inescapable luggage in this endeavor, an issue that would appear to be fairly generic in any attempt to create a covariant theory of MOND.
 Interestingly, Milgrom's BIMOND (Milgrom 2009) does seem to be an exception. 

\subsection{Living without Birkhoff?}

The final point of concern that will be raised here is that MOND, and covariant theories with MOND as a limit
generically lack Birkhoff't theorem.

In Newtonian gravity, the inverse square law guarantees that the gravitational flux through
an enclosed surface is conserved, this is GaussÕs law. 
In General Relativity  BirkhoffÕs Theorem (BT) plays the same role.
Two particularly important aspects of BT are that: (a) outside a spherically symmetric mass
distribution, the metric depends on the distribution of the matter density only at
second order in the potential, due to the role of binding
energy as a source of gravity (and that serves only to change the mass of the system); 
and (b) a spherically symmetric shell of mass has no effect on the metric in its interior.
We depend on these properties to allow us to calculate almost any
gravitational fields without knowing details of the distribution
of matter all over the Universe. Indeed many
calculations in gravity would become undoable either in
principle or practice, and many others would become
enormously more difficult. 

In general, modifications to GR tend to violate BT (Satz 2005; Moffat 2007).
And yet, almost all calculations of the implications of those modifications
rely on BT, or on other congenial averaging properties of GR that 
seem unexpected in a non-linear theory, and that one should not 
therefore assume persist in modifications of GR.  After all, even in 
GR the averaging problem remains unresolved  -- to what extent is the average of the Einstein
tensor in an inhomogeneous universe the solution of the Einstein equations 
with the spatially averaged stress-energy as a source?   

It is important to  appreciate that BT, or more specifically, the fact that
a spherically symmetric shell exerts no gravitational field on its interior
is a rather surprising feature of GR.  In Newtonian physics, we can understand
that for a mass at a point in the interior of the shell the pull from a spherical cap on one
side is perfectly balanced by the pull from the spherical cap of the same angular
size on the opposite side.  This is true even for an off-center location. 
The same is true only for a test mass in GR. A real off-center mass distorts 
the geometry so that the shell is no longer spherical.  This leads to a force
on the mass toward the center!  However the magnitude of the force is
suppressed both by the ratio of the point mass to the shell mass, 
and by the ratio of the Schwarzschild radius of the shell to its actual radius.
Thus, for most cosmologically relevant situations, the corrections to the Newtonian
intuition are small.    

Now MOND too has a Birkhoff-like theorem.  In the interior of a spherical shell,
\begin{equation}\label{eqn:ModPoisson}
\nabla \cdot [\mu (  \abs{\vec{\del}\Phi}/a_0   )  \nabla \Phi ]=0  \rightarrow \nabla \Phi = 0 \,!
\end{equation}
The problem is that as soon as an off-center mass is introduced into the interior
of the shell, the theorem does not apply, and, unlike in GR, the corrections are no longer small (Dai 2008).
The reason is that it was not actually BT that protected us in GR, but rather the inverse-square-law.
Because the whole point of MOND is that gravity falls off more slowly than  $r^{-2}$,
breaking the spherical symmetry immediately leads to sizable forces in the interior
of a shell.

Cosmologically and astrophysically, above the scale of a single star (and even then),
we almost never precisely satisfy the requirement of spherical symmetry.  Galaxies
are not spherical.   They are embedded in groups or clusters that are not spherically
symmetric.  Those clusters are themselves in environments that break spherical symmetry.
In the context of perturbation theory, the question may be phrased in terms of unexpected
mode-mode coupling.

The question then is how large are the corrections to the conclusions that we draw
through the false application of our inverse-square-law-based intuition to systems
with a more long-range force.    The answer to that question is still not clear.
Some progress has been made on simplified systems -- sets of spherical shells, shells with off-center
point masses, ....  For example (Dai 2008), we know that interior point masses that are ${\cal O}(10\%)$ of 
the mass of the surrounding shell can experience "anomalous" forces of a few percent
of the MOND acceleration.    On the other hand, it has also been shown (Matsuo 2009) that
a spherical shell screens the effects of a perturbation on a shell exterior to it, reducing
the field to first order in the perturbation (${\cal O}(\epsilon)$) to ${\cal O}(\epsilon^{1/2})$.
More work clearly remains to be done to provide confident answers for realistic situations to
questions like -- can we smooth the mass density distribution of a collection of stars and gas 
in order to calculate the local field at some point interior to that distribution, and if so,
what is the level of accuracy that that approximation affords?   How does the answer change
if it is a collection of galaxies and gas in a cluster? or clusters in the universe?  
Living with out the $1/r^2$ force law is a dangerous thing to do, especially in a bumpy
universe.  The universe, or it subsystems, may simply not be accurately computable.

\section {Summary}

A model based solely on General Relativity and the Standard Model of particle physics
fails to account for numerous phenomena on scales from the size of our solar system up
to the size of the observed universe.   These include the kinematics of galaxies and clusters
and the thermodynamics of their gas,  the lensing of light by these structures, and their
very existence so soon after the big bang.   On smaller scales, the Pioneer satellites 
appear to have an anomalous sunward component to their acceleration (although this
may or may not have finally been attributed to anisotropic thermal radiation from the satellites).
On larger scales, the universe is unreasonably homogeneous, isotropic and flat, unexpectedly
old given its rate of expansion,  expanding surprisingly rapidly given the low abundance of matter,
expanding at an accelerating rate, and distinctly  lacking in angular correlations on the largest
angular scales.    The dimensionful scales (lengths, times, accelerations) 
at which of many of these anomalies appear seem to be similar: $H_0$ times appropriate powers
of $c$.

   The concordance model of the universe addresses many of these puzzles through the introduction 
   of three separate dark sectors -- dark matter, dark energy and the inflaton.   It is a very successful
   model (though mysteries remain), but its arbitrariness motivate us to consider the possibility that
   it is not, or not just, that the stress energy content of the Standard Model  is incomplete but that
   our theory of gravity needs modification.   The most challenging sector of the concordance model
   to replace with modified gravity is dark matter, because it ties together with reasonable success
   many of the problems.  Modified Newtonian Dynamics provides a phenomenological framework
   within which to address many of these questions.  It has particular   successes of its own, such as 
   explaining the relationship between galaxy luminosities and kinematics.  However it also forces
   us to give up rather useful features of GR like Birkhoff's theorem.
   
   Covariant theories have been
   built that approach MOND in the appropriate limits.   These theories may address some of the challenges
   of MOND -- such as the failure to explain clusters without the introduction of hot dark matter, 
   and the failure to grow large scale structures.  However they have their own challenges.  
   
   The program of exploring modified gravity continues to be a worthwhile one, and most excitedly 
   modifying gravity on cosmological scales to explain cosmic acceleration seems both doable and testable.
   Replacing dark matter with modified gravity seems a more difficult road to travel despite the
   construction of covariant theories with MOND limits.  At least we discover that they only seem to work 
   if they  bring back dark matter in the form of dark fields.   That may be how the Universe works, but
   it isn't what the doctor (Milgrom) ordered.

\section {Acknowledgements}
GDS thanks his collaborators D. Dai, P. Ferreira, D. Jacobs, I Maor, R. Matsuo  and T. Zlosnik
for numerous and extensive conversations regarding many of the issues discussed above.  He
apologizes in advance to his many other colleagues whose work on some of these issues he
may have failed to adequately cite, and looks forward to receiving their assistance in correcting
these shortcomings.  GDS is supported by grants from the US DOE and NASA ATP.


\smallskip{}

\end{document}